\documentclass[twocolumn,aps,showpacs]{revtex4}
\usepackage{graphicx}
\usepackage{amsmath}
\usepackage{amsfonts}
\usepackage{amssymb}

\begin{document}

\title{\textbf{Optimization and self-organized criticality in a magnetic system}}
\author{Roberto N. Onody}
\email{onody@if.sc.usp.br}
\author{Paulo A. de Castro}
\email{pac@if.sc.usp.br}
\affiliation{Departamento de F\'{\i}sica e Inform\'atica,
Instituto de F\'{\i}sica de S\~ao Carlos, \\
Universidade de S\~ao Paulo, C.P.369, 13560-970 S\~ao Carlos-SP,
Brazil}

\begin{abstract}
We propose a kind of Bak-Sneppen dynamics as a general
optimization technique to treat magnetic systems. The resulting
dynamics shows self-organized criticality with power law scaling
of the spatial and temporal correlations. An alternative method of
the extremal optimization is also analyzed here. We provided a
numerical confirmation that, for any possible value of its free
parameter $\tau$, the extremal optimization dynamics exhibits a
non-critical behavior with an infinite spatial range and
exponential decay of the avalanches. Using the chiral clock model
as our test system, we compare the efficiency of the two dynamics
with regard to their abilities to find the system's ground state.


\end{abstract}
\maketitle

\section{Introduction}

Nature which is composed of dead and living things is a system far
out of equilibrium. This can be evidenced by measuring nature's
evolution with the appropriate scales of space and time. In
particular, the dynamics of the living things is basically
governed by the theory of natural selection first proposed by
Charles Darwin \cite{cd}. This theory is certainly one of greatest
scientific achievements of the nineteenth century. Very briefly,
this theory simply says that evolution occurs not by breeding
species best adapted to their environment but by driven to
extinction those less or poorly adapted. This mechanism leads to
the emergence of highly specialized structures. If we also
consider the astonishing variability of the species, we then can
say that nature is a complex system. Indeed, for all we know,
nature operates \textit{at} the self-organized critical state
\cite{bak}.

One of the most fundamental characteristics of a system with
self-organized criticality (SOC) is to exhibit a stationary state
with a long-range power law decay of the spatial and temporal
correlations \cite{btw}. Power law is a very abundant behavior
found either in natural phenomena such as the light emitted from
quasars, the earthquake's intensity, the water level of the Nile
or as a direct result of human activities like the distribution of
cities by size, the repetition of words in the Bible and in
traffic jams.

Self-organized critical systems evolve to the complex critical
state without the interference of any external agent. Differently
from what happens in the usual critical phase transition, in SOC
there is not a tuning parameter. The prototypical example of SOC
is a pile of sand \cite{btw}. Another common property of SOC is
that the self-organization state only takes place after a very
long transient period. Last but not least, a minor change in the
system can cause colossal instabilities called avalanches.
Intermittent bursts of activity separating long periods of
quiescence is called punctuated equilibrium. Gould has conjectured
that the biological evolution itself shows, in fact, this kind of
equilibrium \cite{gou}.

Now, if nature can really be found in a critical self-organized
state, is this state optimal or extremal in some sense?. Through
the systematic rejection of the worst, the Darwinian evolutionary
theory is a born optimization structure. By adding interactions
and competition between the species we will witness the emergence
of co-evolution.

One model specially tailored to represent the co-evolutionary
activities of the species is the Bak-Sneppen model \cite{bs}. In
this model, each species is located on the sites of a lattice and
has associated a fitness value between 0 and 1 (randomly drawn
from an uniform distribution). At each time step, the species with
the smallest associated value as well as its nearest neighbors are
selected to replace their fitness with new random numbers. In one
dimension, after a long transient time, almost all species have
fitness larger than the critical value 0.67 \cite{bs}.

Recently, inspired by natural processes, some heuristic
optimization techniques have been proposed: genetic algorithms
\cite{hol}, simulated annealing \cite{kirkpa} and extremal
optimization \cite{boe}. The extremal optimization (EO) method
seems to be the most efficient of them since it brings the system
faster and closer to its ground-state \cite{boe}. In a few words,
this method consists of the following rules: i) a discrete
dynamical variable $S_{i}$ (initially chosen at random) is
associated to each site $i$ of a lattice with $N$ points; ii) a
fitness $ \lambda_{i}$ is attributed to that site using some given
prescription; iii) all lattice sites are then increasingly ranked
according to their fitness (the site with the worst fitness is of
rank 1); iv) a site of rank $k$ ($ 1 \le k \le N$) is selected
with probability $P(k) \propto k^{- \tau}$ ($\tau$ is an arbitrary
real positive number) and its dynamical variable $S_{i}$ is
changed to $S_{i}^{'}$ ($S_{i}^{'} \ne S_{i} $, compulsorily); v)
repeat at step iii) as long as desired.

In this paper, a kind of Bak-Sneppen dynamics is applied to a spin
system which possesses a complex ground state structure in one
dimension. The \textit{discrete fitness variability} is identified
as being responsible for the \textit{absence} of a critical
self-organized state. By introducing a noise in the spin
configuration space, we show that the system can now reach a
critical self-organized state with power law correlations. This
simple trick can easily be extended to other types of discrete
systems. On the other hand, through an analysis of the spatial and
temporal correlations, we provided a numerical confirmation that
the extremal optimization method is not a SOC dynamics but a
non-critical algorithm with an infinite spatial range and an
exponential time decay of the avalanches. The two dynamics: the
Bak-Sneppen with noise (BSN) and the extremal optimization (EO)
are then investigated with respect to their efficiency to minimize
energy. There is an optimal interval of the parameter $\tau$ where
EO is superior to BSN.

\section{The p-states chiral clock model}

To explain our main ideas, we have chosen the one dimensional
p-states chiral clock model ($CC_{p}$) \cite{ost,huse} as our
experimental system. In higher dimensions, this system exhibits a
complex phase transition diagram with commensurate and
incommensurate phases. In one dimension, the competition between
the applied magnetic field, which tries to align the spins, and
the chirality, which tries to flip them, gives rise to a rich
ground-state diagram in the space of the parameters. The
hamiltonian is given by

%------------------------------------ equation 1 ------------------------------
\begin{eqnarray}
H = \sum_{i=1}^{N}\{1-\cos(2 \pi (
\frac{S_{i+1}-S_{i}}{p}-\Delta)) + \notag
\\
h [1-\cos(2 \pi (\frac{S_{i}-1}{p}))]\} = -\sum_{i=1}^{N}
\lambda_{i}
\end{eqnarray}
%------------------------------------------------------------------------------
where $p$ is a positive integer number, $S_{i}=1,..,p$ is the spin variable
at the site $i$ (with periodic boundary conditions applied) and
$h$ and $\Delta$ are the magnetic field and chirality, respectively.
By symmetry arguments,
$h$ and $\Delta$ may be restricted to the intervals $[0,\infty]$ and $[0,1/2]$.

\begin{figure}[htbp!]
\begin{center}
\includegraphics[width=8cm]{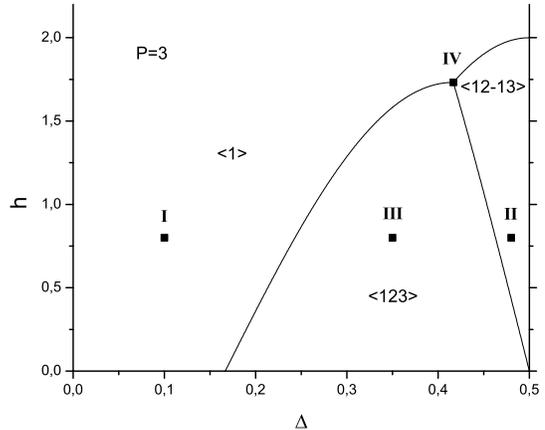}
\end{center}
\caption{The magnetic field $h$ versus the chirality $\Delta$ for
the ground state diagram of the $p=3$ chiral clock model. The
regions $<1>$, $<12-13>$ and $<123>$ have periods 1, 2 and 3,
respectively. The marked points $I$, $II $, $III$ and $IV$ were
tested for the universality of the BSN dynamics.}%
\label{Figure1}%
\end{figure}

For $p=3$, the chiral clock model has 3 regions with different
ground-states (see Fig.1). We denote by $<1>$, $<12-13>$ and
$<123>$ the regions with periods 1, 2 and 3, respectively. The
numbers correspond directly to the sequence of the spin states.
The region $<12-13>$ is super degenerate, i. e., the spin state
sequence 12 may be equally followed by 12 or 13. For a finite
ring, this means that the degenerescence of this region grows
exponentially with the lattice size.

\section{Searching the ground-state with a Bak-Sneppen dynamics}

We choose the following form for the fitness $\lambda_{i}$

%------------------------------------ equation 2 ------------------------------
\begin{eqnarray}
- \lambda_{i} & = & \frac{1}{2} [1-\cos(2 \pi (
\frac{S_{i+1}-S_{i}}{p} - \Delta ))] + \notag \\
& & \frac{1}{2} [1-\cos(2 \pi ( \frac{S_{i}-S_{i-1}}{p} - \Delta
))] + \notag \\
& & \frac{1}{3} h [1-\cos(2 \pi ( \frac{S_{i+1}-1}{p}))] + \notag \\
& & \frac{1}{3} h [1-\cos(2 \pi ( \frac{S_{i}-1}{p}))] + \notag \\
& & \frac{1}{3} h [1-\cos(2 \pi ( \frac{S_{i-1}-1}{p}))]
\end{eqnarray}
%------------------------------------------------------------------------------

There is a certain art and arbitrariness to select the fitness
form. For instance, one could equally well have chosen a
pre-factor 1/4 for the magnetic fields at the sites $i+1$ and
$i-1$ and 1/2 for the site $i$. Different fitness forms may lead
to quite different values of some physical quantities determined
by the dynamics, e. g., the average energy in the stationary
state. The fitness defined by Equation (2) is the form with the
highest symmetry - the magnetic field is equally distributed
between the sites $i-1$, $i$ and $i+1$ as well as is the chirality
between the pair of sites $[i,i-1]$ and $[i+1,i]$. It represents
the best result we got for the average energy in a bunch of
trials.

Let us now describe our procedure. Initially, to each site $i$ of
the ring, we attribute by chance one of the $p$ possible values of
the spin variable $S_{i}$. Using equation (2), we calculate all
the corresponding fitness $\lambda_{i}$ and find the smaller
(worst) one $\lambda_{j}$. New spin variables are then randomly
assigned to the sites $j$, $j+1$ and $j-1$. So although the
dynamics involves 3 spins, due to our fitness definition, it
affects (changes) 5 consecutive fitness (from the site $j-2$ to
$j+2$). Observe that here the Bak-Sneppen dynamics is being
applied to the \textit{spin configuration} space not to the
\textit{fitness} space itself (as it was originally done in the
evolutionary context \cite{bs}).  In the final step, a new site
with the smallest fitness is searched and the whole process
continues as long as we wish. However, even for a modest lattice
size, this dynamics is \textit{hampered} by the discrete fitness
variability. Let us explain why. If $p=3$ ($h$ and $\Delta$ fixed)
there are at most 27 possible values of the fitness (besides, some
of these may be degenerate). This means that, for a lattice with a
reasonable size, we will generally find not just \textit{one} but
a \textit{large} number of sites with exactly the
\textit{same}(smallest) fitness value. The simplest solution to
the problem seems to put all those sites in a list and to choose
one of them at random. We will call this dynamics as the
Bak-Sneppen with draw (BSD). As we shall see later, neither from
the point of view of the spatial correlations nor from the time
correlations is this dynamics a true SOC.

\section{The noise}
It is the discrete variability of the fitness (or, equivalently,
of the spin variable) which precludes the establishment of a
self-organized critical state, so we searched a simple way to
solve the problem. Suppose that a noise $\eta$ is added to the
original spin variable $S_{i}$ (an integer in the interval
$[1,p]$), i. e.

%------------------------------------ equation 3 ------------------------------
\begin{equation}
S^{'}_{i} = S_{i} + \eta (1-2r)
\end{equation}
%------------------------------------------------------------------------------
where $r \in [0,1]$ is a random number generated from a flat
distribution. The new spin variable $S^{'}_{i}$ as well as the
corresponding fitness have now the desired continuous
characteristic. By choosing the noise $\eta$ sufficiently small
the relevant physical properties (like the energy per site or the
magnetization per site) will not be affected. In this paper we set
$\eta = 10^{-3}$. Using double numerical precision, we varied $
\eta $ from $ 10^{-12}$ up to $10^{-3}$ with no important
difference for the physical quantities. With this trick, any
previously discrete fitness $\lambda$ is turned into a continuous
variable inside some interval $\Delta \lambda$ controlled by the
noise $ \eta $, the chirality $ \Delta $ and the magnetic field
$h$. We will call this dynamics as the Bak-Sneppen with noise
(BSN). Our next objective is to compare the 3 dynamics EO, BSD and
BSN with respect to their efficiencies.

In order to guarantee that the stationary state had already been
reached and to get good averages for the physical quantities, a
huge amount of computation was performed. Using the same fitness
definition (equation (2)) for all the three algorithms BSD, BSN
and EO, we simulated each one of them over $2$ $10^{9}$ runs on a
ring of 4001 sites. Discarding the first $ 2$ $10^{8} $ runs ($ 10
\% $ of the total) as the transient time, averages were then taken
over the remaining steps. The results for the energy histograms
are shown in Figure 2.

\begin{figure}[htbp!]
\begin{center}
\includegraphics[width=8cm]{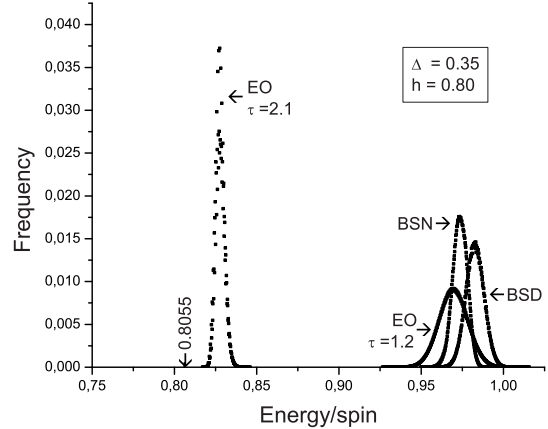}
\end{center}
\caption{Energy per spin histograms for the EO (with $\protect\tau
= 2.1$ and $1.2$), BSN and BSD dynamics. They were determined at
the point $III$ of the Figure 1. The exact ground state energy per
spin is 0.8055.}%
\label{Figure2}%
\end{figure}

The simulations were done at the point $III$ of the region $
<123>$ (see Figure 1) where the ground state energy per spin is
equal to $0.8055$. The best performance was obtained by the EO
algorithm with $ \tau = 2.1$, followed by the EO with $ \tau =
1.2$, BSN and BSD. Their respective mean energies were $0.828 \pm
0.003$, $ 0.969 \pm 0.009$, $ 0.973 \pm 0.004$ and $ 0.983 \pm
0.005$.

\section{The spatial and temporal distributions}

To study the spatial correlation, we measured the distribution $
D(x) $, of the distance $x$ between two subsequent activated sites
\cite{bs}. Recall that for the three algorithms we activate a site
$i$ by changing its spin state $S_{i}$ and the corresponding
fitness $ \lambda_{i}$. We plotted $D(x)$ in the Figure 3.

\begin{figure}[htbp!]
\begin{center}
\includegraphics[width=8cm]{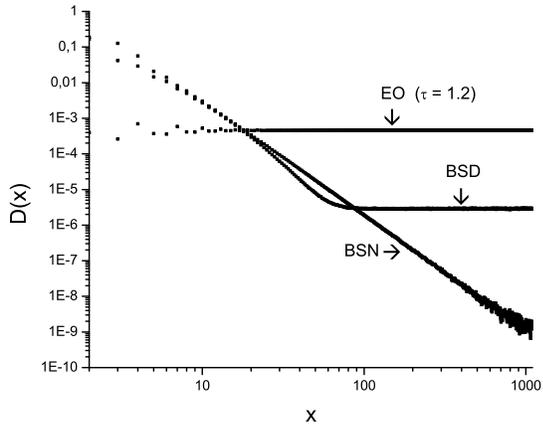}
\end{center}
\caption{The probability distribution $D(x)$ versus the distance
$x$ between two subsequent activated sites. The algorithm EO shows
an infinite spatial range while the BSN algorithm has a power
law scaling.}%
\label{Figure3}%
\end{figure}

Clearly, EO is an algorithm with an \textit{infinite spatial
range} - it doesn't matter how far one site is of the other, the
activation probability is constant. On the other hand, the BSN
algorithm exhibits a \textit{power law} dependence $ D(x) \sim
x^{-3.19 \pm 0.02}$ with the exponent being compatible with that
of the Bak-Sneppen model ($ 3.23 \pm 0.02$ \cite{maya}). Moreover,
we got at the points $I$, $II$, $III$ and $IV$ (see Figure 1) the
same exponent value, indicating that the universality principle
holds in the space of the parameters $h$ and $\Delta$. Finally,
from the Figure 3 we conclude that the BSD algorithm is of a mixed
kind, having the BSN behavior for small distances and the EO for
large distances.

In the stationary regime, the BSN dynamics shows an intermittent
time evolution with long periods of passivity interrupted by
sudden bursts of activity, i. e., it exhibits a \textit{punctuated
equilibrium}. This abrupt change of activity is called an
avalanche. We say that an avalanche is happening if there exist
one site of the ring whose fitness is smaller than a certain
threshold $\lambda_{c}$. The size $A$ of an avalanche is defined
as the number of subsequent time steps with at least one fitness
below that threshold. For each one of the three dynamics we
calculated the probability distribution of the avalanches $P(A)$.
The results are illustrated in the Figure 4.

\begin{figure}[htbp!]
\begin{center}
\includegraphics[width=8cm]{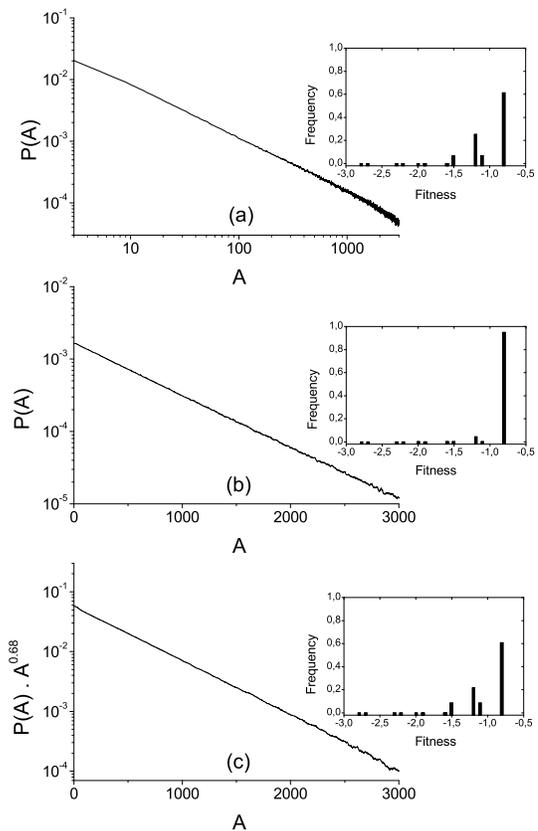}
\end{center}
\caption{Plot of the probability distribution of the avalanches
$P(A)$ versus the avalanches size $A$ for the BSN, EO and BSD
algorithms in (a), (b) and (c), respectively. The insets
correspond to the
fitness histograms evaluated at the time step $2$ $10^{8}$.}%
\label{Figure4}%
\end{figure}

The Figures 4(a), 4(b) and 4(c) correspond to the BSN, EO and BSD
algorithms, respectively. Before we analyze $P(A)$, let us first
discuss how the critical thresholds were obtained. At the time
step $2$ $10^{8}$ (the end of the transient period) and with the
magnetic field and chirality fixed at the values $0.8$ and $0.35$,
respectively, we determined the fitness histograms of the whole
ring. They are shown as insets of the Figure 4. Since in the EO
and the BSD algorithms there are only some few possible values for
the fitness $\lambda$ (a maximum of 27 when $p=3$ and many of them
are degenerate), we may vary the (discrete) $\lambda$ values until
the avalanches are founded. Remember that if $\lambda <
\lambda_{c}$ there will be no avalanches while if $\lambda
> \lambda_{c}$ only one (infinite) avalanche is present. Due to
the discreteness values of $\lambda$, it is very improbable to
find the system in a sub or supercritical regime. We found
$\lambda_{c} = -0.8055$ and $\lambda_{c} = -1.5061$ for the EO and
BSD dynamics, respectively. For the BSN algorithm, however, the
fitness is continuous. There is a fitness fluctuation $\Delta
\lambda$ due to the presence of the noise $\eta$. In our case, the
maximum value of the relative fluctuation $\Delta \lambda$ /
$\lambda$ is $1.86$ $10^{-3}$.  To find the critical threshold, we
wrote a program which varies the fitness values (inside the range
$\Delta \lambda$) until the resulting avalanches have sizes
between 500 and 2000 (which seems reasonable to avoid the sub and
supercritical regimes). We determined $\lambda_{c} = -1.5059$.

We can now go back to the probability distribution of the
avalanches $P(A)$. They were all calculated at their respective
thresholds $\lambda_{c}$. The EO algorithm (Figure 4(b)) shows an
\textit{exponential} decay with $A$. The BSN (Figure 4(a)), on the
other hand, has a \textit{power law} scaling $P(A) \sim A^{-0.87
\pm 0.01} $. This value is somewhat discrepant from that of the
evolutionary Bak-Sneppen model ($1.07 \pm 0.01$ \cite{maya}), but
we should remember that, for our system, the exact determination
of the critical threshold position is much more difficult and one
can easily be found in a sub or supercritical regime. For the BSD
algorithm (Figure 4(c)), we found once more a mixed behavior
described by $ P(A) \sim A^{-0.68} e^{-0.002 A}$.

\section{Conclusions}

Based on the Bak-Sneppen ideas, we proposed an optimization
algorithm whose dynamics, in the steady state regime, exhibits
self-organized criticality. It has been successfully applied to
find the ground states of the clock chiral model and it can be
easily extended to any other magnetic system. From the three
dynamics studied in this paper, we conclude that only the BSN
algorithm conducts the system to a self-organized critical state.
The essential point was to recognize that the discrete fitness
values, present in the EO and BSD algorithms, preclude the system
to have a power law decay of the spatial correlation. To overcome
this problem we introduced a small noise in the spin configuration
space. Another important conclusion we arrived is that to possess
SOC characteristics does not guarantee a dynamics to be optimal.
This statement was proved here for the EO algorithm. The EO
algorithm, which is not SOC since it has an infinite spatial range
and the avalanches decay exponentially, has a peak of the energy
histogram (with $ \tau = 2.1$) which is only $ 2.7 \% $ further
from the exact ground state value. This is a stunning result which
cannot be beaten by the BSN dynamics. At $\tau = 1.2$, the EO and
BSN are almost equivalent. On the other hand, the BSD algorithm is
of a mixed type. It can be obtained by taking the limit $\eta
\rightarrow 0$ in the BSN.

Finally, it is important to observe that, contrary to what happens
with BSN, the EO algorithm has one arbitrary and free parameter
$\tau$. If $ \tau \rightarrow 0$, EO becomes a random walk and if
$ \tau \rightarrow \infty$, the system may stack in a dead end. In
both limits, the EO efficiency is completely spoiled and an
optimal value of $\tau$ should be found in somewhere between. In
the BSN dynamics there is no tuning parameter. Using the ideas
described here, we wish to develop a SOC optimization algorithm to
some classical combinatorial problems like the bipartition of
graphs.

This work was supported in part by the CNPq (Conselho Nacional de
Desenvolvimento Cient\'{\i}fico e
Tecnol\'ogico) and by the FAPESP (Funda\c c\~ao de Amparo a Pesquisa do
Estado de S\~ao Paulo).

\end{document}